\documentstyle[12pt]{article}
\def\be{\begin{equation}}
\def\ee{\end{equation}}
\def\bea{\begin{eqnarray}}
\def\eea{\end{eqnarray}}
\def\a{\alpha}
\def\b{\beta}

\def\e{\epsilon}

\author{Hans-J\"urgen Schmidt}

\title{Comment on ``Stability of the semiclassical Einstein equation"}

\date{}
\begin{document}
\maketitle

\centerline{Universit\"at Potsdam, Institut f\"ur Mathematik, Am
Neuen Palais 10} 
 \centerline{D-14469~Potsdam, Germany,  E-mail:
 hjschmi@rz.uni-potsdam.de}

\begin{abstract}
\noindent
Some mathematical errors of 
the paper commented upon are corrected. 

\medskip
\noindent 
PACS number(s): 04.50.~h, 04.20.Jb, 04.60.Ds
\end{abstract}

\bigskip

The calculations of Ref.  [1] are analyzed; the results apply 
analogously also to Ref.  [2]. The reason we do it now is 
that Ref. [3] uses the statements in Ref. [1] as motivation for 
further development. In units where $8 \pi  G = c = 1$
 we consider
 the Lagrangian
\be
L = R/2 - l^2 \, R^2\, / 12 + \a C_{ijkl}C^{ijkl} \, , 
\ee
where $R$ is the curvature scalar and 
$C_{ijkl}$  denotes the Weyl tensor. 
It holds that   
$ {\rm sgn}(\e) = {\rm  sgn}(l^2)
 = {\rm sgn}(- \mu) = {\rm  sgn}(\b) \, , 
$ 
where $\e$ is from [1,2], $l^2$ from Eq. (1), $\mu$ from
 [4], and $\b$ from [5]. 
All authors agree that for $l^2 < 0$ unwanted instabilities 
appear, so we do not further discuss this case. 
The term with a in Eq. (1) is discussed, 
e.g., in [5] (this
 is the first paper of Ref. [11] in [1]): See Eqs. (4) and (17) of [5]; 
the text around Eq. (35) of [5] makes it dear that the exponentially 
growing solutions come from the term with $\a$, and additional 
such  solutions appear only for $l^2<0$.

\bigskip

However, for the Friemann  models 
considered in [1,2], the term with $\a$ identically vanishes because
 of the conformal flatness of the models. In the Starobinsky model, 
one has $\a = 0$  from the beginning, so that the Horowitz instabilities [5] 
do not appear even for general space--times. So the phrase ``This
 may seen to be in contradiction to 
Horowitz's result" at the end of p. 315 of [1] should 
be replaced by ``This is not in contradiction with 
Horowitz's result." Let 
us put $\a =0$ in the following.

\bigskip

In [1,2] three further kinds of terms are considered. This 
refers to the terms with coefficients  $P_i$, $i = 1,2,3$.
 Reference 
 [1] on p. 318 and Ref. [2], p. 2218 has ``For most cases of interest, 
 $P_1$ is much less than $1/G$, and $P_2$ and $P_3$
 are much less than $\e$'' and, 
on p. 2219,  ``have this \dots 
situation independent of the \dots  values of the
 $P_i$'s''. It agrees 
with my own (unpublished) calculations that for the 
qualitative behavior of the solutions the terms with the $P_i$'s do 
not play a role. So we set them to zero without loss of generality.

\bigskip

The phrase  ``these oscillations are damped by the back reaction 
of particle production" at the end of p. 315 of [1] is  misleading 
because they are already damped without considering back 
reactions (cf. [4]), and the back reaction only intensifies the damping.

\bigskip

 Let us now come to the main point: the qualitative
 behavior of the spatially flat Friedmann solutions of the
 field equation of fourth order following from the 
 Lagrangian (1) with $\a = 0$ and $l^2 > 0$. To ease the comparison 
between Refs. [1] and [4], one should notice that, in
[4], the time direction is chosen such that the Universe
  expands (this excludes a negative Hubble parameter $H$), 
 so [1] with $H < 0$ has to be compared with an inverted 
time direction and $H> 0$ in [4].

\bigskip

 The Appendix uses an ansatz (Al); it should be noted 
 that it does not converge. Consequently, in the phrase
``$a(t) \to  a_0 \, t^{1/2}$ at
 late time" (p. 325 in [1]) the exponent
     1/2 has to be replaced by 2/3. (The reason is that, while
 expanding, the influence of the radiation becomes less
     and less in comparison with the curvature squared term 
          until it is negligible.)

\bigskip

 Equation (3.1) of [1] contains the Hubble parameter $H$
in the denominator; so it is  clear that $H \to 0$ represents a 
     singular point of the differential equation; consequently,
     the usual perturbation technique as Suen applies it need
     not give correct results. For example, on p. 318 (right 
column) he writes ``For $H_i < 0$ \dots we have a catastrophic
          collapse to a singularity." A counterexample is Eq. (18)
          of [4], which shows that there is a solution that remains
          regular during infinite time. All other solutions indeed
          have this catastrophic collapse (in the notation of [4] it
          lies in the past),  but it represents nothing but the big
          bang of general relativity. So in Ref. [2], p. 2220, left
          column, the statement ``the SCE theory would be in serious
  trouble, unless $\e$ is exactly zero" gives the impression
          that $\e = 0$ would diminish these troubles (SCE denotes
          semiclassical Einstein). However, one has to say that the
          fourth-order theory following from Eq. (1) with $\a = 0$
          and $l^2 > 0$ is not more unstable than Einstein's theory
          itself. To be precise, one has to add that additional 
 instabilities can occur only for $R \ge 3/l^2$. But this is
 not a  real restriction because $l$ is  microscopically small and the
   
 interesting inflationary phase has a negative curvature
          scalar.

\bigskip

           Concerning  the case $H_i > 0$, p. 318, right column,
          Ref. [1] reads ``However, it is physically unacceptable
  \dots'' This is not the case; on the contrary (see [4]) for
          this case all solutions are well behaved up to infinite time.

\bigskip

\noindent
[1] W.-M. Suen, Phys. Rev. D {\bf 40}, 315 (1989).

\noindent 
[2] W.-M. Suen, Phys. Rev. Lett. {\bf 62}, 2217 (1989). 

\noindent
[3] J. Simon, Phys. Rev. D {\bf 45}, 1953 (1992).

\noindent
[4] V. M\"uller and H.-J. Schmidt, Gen. Relativ. Gravit. 
{\bf 17}, 769 (1985).

\noindent
[5] G. T. Horowitz, Phys. Rev. D {\bf 21}, 1445 (1980).

\bigskip

(Received 15 November 1993)

\bigskip

\noindent
{\small {This is a  reprint of 
 Phys.  Rev.  D, Vol. 50, Nr. 8,  1994, page 5452. 
}}
\end{document}